\documentclass[twocolumn,showpacs]{revtex4}

\usepackage{graphicx}%
\usepackage{dcolumn}
\usepackage{amsmath}

\makeatletter
\def\btt#1{\texttt{\@backslashchar#1}}%
\DeclareRobustCommand\bblash{\btt{\@backslashchar}}%
\makeatother

\begin{document}

\title[Short Title]{The Tachyon Inflationary Models with Exact Mode Functions }

\author{Xin-zhou Li}
 \email{kychz@shtu.edu.cn}
\author{ Xiang-hua Zhai}%

\affiliation{%
Shanghai United Center for Astrophysics,Shanghai Normal
University, Shanghai 200234 , China
}%

\date{\today}

\begin{abstract}
We show two analytical solutions of the tachyon inflation for
which the spectrum of curvature (density) perturbations can be
calculated exactly to linear order, ignoring both gravity and the
self-interactions of the tachyon field . The main feature of these
solutions is that the spectral indices are independent with scale.
\end{abstract}

\pacs{04.20.Jb, 98.80.Cq}

\maketitle

\section{Introduction}
\label{sec:level1}

The inflationary scenario[1]provides a mechanism to produce the
primordial fluctuations of spacetime and matter, which lead to the
CMBR(cosmic microwave background radiation)anisotropies and to the
large scale structure[2]. In standard inflationary models[3], the
physics lies in the inflation potential. The underlying dynamics
is simply that of a single scalar field rolling in the potential.
The scenario is generically referred to as chaotic inflation in
reference to its choice of initial conditions. This picture is
widely favored because of its simplicity and has received by far
the most attention to date. Some potentials that give the correct
inflationary properties have been proposed[4]in the past two
decades. Recently, pioneered by Sen[5], the study of non-BPS
objects such as non-BPS branes, brane-antibrane configuration or
space-like branes[6] has been attracting physical interests in
string theory. Sen showed that classical decay of unstable D-brane
in string theories produces pressureless gas with non-zero energy
density[7]. Gibbons took into account coupling to gravitational
field by adding an Einstein-Hilbert term to the effective action
of the tachyon on a brane, and initiated a study of "tachyon
cosmology"[8]. The basic idea is that the usual open string vacuum
is unstable but there exists a stable vacuum with zero energy
density. There is evidence that this state is associated with the
condensation of electric flux tubes of closed string[7]. These
flux tubes described successfully using an effective Born-Infeld
action[9]. The string theory motivated tachyon inflation[10] and
quintessence[11]have been discussed. These investigation are based
on the slow-roll approximation so that they are rather incomplete.
It is obvious that tachyon inflation might occur in a much broader
context and we should discriminate between generic predictions of
inflation and predictions of a specific scenario.

For any model of inflation, the scalar power spectrum can be
expressed as a series[12]. One often rely on either on
approximations or on numerical mode-by mode integration for
predicting the coefficients of the series. However, the accuracy
of any approximation scheme should be controlled which means that
the errors should quantified. This can only be done by comparing
the approximate power spectrum to an exact one[13]. Therefore, it
is very interesting to find the exact solutions of mode equation
for the tachyon inflation.

\section{The equations of motion}

We consider spatially flat FRW line element given by:

\begin{eqnarray}
ds^{2}&=&dt^{2}-a^{2}\left(t\right)\left(dx^{2}+dy^{2}+dz^{2}\right)\nonumber\\
&=&a^{2}\left(\tau\right)\left[d\tau^{2}-\left(dx^{2}+dy^{2}+dz^{2}\right)\right]\hspace{3cm}
\end{eqnarray}

\noindent where $\tau$ is the conformal time, with $dt=ad\tau$. As
shown by Sen[7], a rolling tachyon condensate in either bosonic or
supersymmetric string theory can be described by a fluid which in
the homogeneous limit has energy density and pressure as follows

\begin{eqnarray}
\rho &=&\frac{V(T)}{\sqrt{1-\dot{T}^2}}\nonumber\\
p&=&-V(T)\sqrt{1-\dot{T}^2}\hspace{3cm}
\end{eqnarray}

\noindent where $T$ and $V(T)$ are the tachyon field and
potential, and an overdot denote a derivative with respect to the
coordinate time $t$. To take the gravitational field into account
the effective lagrangian density in the Born-Infeld-type form is
described by:

\begin{equation}
L=\sqrt{-g}\left(\frac{R}{2\kappa}-V(T)\sqrt{1+g^{\mu\nu}\partial_{\mu}T\partial_{\nu}T}\right)\hspace{3cm}
\end{equation}

\noindent where $\kappa=8\pi G=M_p^{-2}$. For a spatially
homogeneous tachyon field $T$, we have the equation of motion

\begin{equation}
\ddot{T}+3H\dot{T}\left(1-\dot{T}^2\right)+\frac{V^{\prime}}{V}\left(1-\dot{T}^2\right)=0\hspace{3cm}
\end{equation}

\noindent which is equivalent to the entropy conservation
equation. Here, the Hubble parameter $H$ is defined as $H\equiv
\dot{a}/a$, and $V^{\prime}=dV/dT$ . If the stress-energy of the
universe is dominated by the tachyon field $T$, the Einstein field
equations for the evolution of the background metric,
$G_{\mu\nu}=\kappa T_{\mu\nu}$, can be written as

\begin{equation}
H^2=\frac{\kappa}{3}\frac{V(T)}{\sqrt{1-\dot{T}^2}}\hspace{3cm}
\end{equation}

\begin{equation}
\frac{\ddot{a}}{a}=H^2+\dot{H}=\frac{\kappa}{3}\frac{V(T)}{\sqrt{1-\dot{T}^2}}\left(1-\frac{3}{2}\dot{T}^2\right)
\hspace{3cm}
\end{equation}

 \noindent From Eqs.(5) and (6) we deduce that

\begin{equation}
\frac{dT}{dt}=-\frac23\frac{H^{\prime}(T)}{H^2(T)}\hspace{3cm}
\end{equation}

\noindent and leading to

\begin{equation}
V^2(T)=\frac9{\kappa^2}H^4-\frac4{\kappa^2}H^{\prime^2}\hspace{3cm}
\end{equation}

\begin{equation}
\frac{a}{a_0}=\exp\left[-\frac32\int^T_{T_0}\frac
{H^3}{H^{\prime}}dT\right]\hspace{3cm}
 \end{equation}

 \begin{equation}
 t=-\frac32\int^T_{T_0}\frac{H^2}{H^{\prime}}dT\hspace{3cm}
 \end{equation}

\noindent where $a_0$ is the initial value of scale factor during
inflation.

Bardeen et.al. [14] have shown that the general form of the metric
for the background and scalar perturbation id given by

\begin{eqnarray}
a^{-2}(\tau)ds^2&=&(1+2A)d\tau^2-2\partial_iBdx^id\tau\nonumber\\
&-&\left[(1-2R)\delta_{ij}+2\partial_i\partial_jE\right]dx^idy^j
\end{eqnarray}

\noindent The intrinsic curvature perturbation $\mathcal{R}$ of
the comoving hypersurfaces can be written as

\begin{equation}
{\mathcal R}=-R-\frac{H}{\dot{T}}\delta{T}\hspace{3cm}
\end{equation}

\noindent during evolution of the universe, where $\delta{T}$
represents the fluctuation of the tachyon field and $\dot{T}$ and
$H$ are determined by the background field equations Eqs.(7)-(8).
On the analogy of discussion for the inflation driven ordinary
scalar field with the standard kinetic term[15], we use the
gauge-invariant potential

\begin{equation}
u=a\left[\delta T+\frac{\dot{T}}{H}R\right]\equiv-z{\mathcal
R}\hspace{3cm}
\end{equation}

\noindent where we have introduced the new variable

\begin{equation}
z\equiv\frac{a\dot{T}}{H}\hspace{3cm}
\end{equation}

The evolution of the scalar perturbations is calculated by the
Einstein action. The first-order perturbation equations of motion
are given by a second-order action. Therefore, the gravitational
and matter sectors are separated and each expanded to second-order
in the perturbations. The action for the matter perturbations can
be determined by expanding the Lagrangian as a Taylor series about
the background equations and integrating by parts. Note that the
inflationary requirement $\ddot{a}>0$ as long as
$\dot{T}^2<\frac23$. In the chaotic scenario, the inflation will
slowly roll down its potential, i.e., $\dot{T}^2\ll \frac23$ and
$H^2\approx \frac{\kappa}3V(T)$. Hence, the tachyon equation of
motion (4) is approximated to the one with the Lagrangian of the
normal nearly quadratic form. Furthermore, we can state a small
parameter
$\sqrt{\epsilon}=\sqrt{\frac{2}{3}}\left(\frac{H\prime}{H^2}\right)$
which suppressed the non-linear contributions.  The action of
linear scalar perturbation of tachyon field is given by

\begin{equation}
{\mathcal{S}}=\frac12\int d\tau d^3{\mathbf{x}}
\left[(\partial_{\tau}u)^2-\delta^{ij}\partial_iu\partial_ju
+\frac{z_{\tau\tau}}{z}u^2\right]\hspace{3cm}
\end{equation}

\noindent Eq.(15) is formally equivalent to the action for a
scalar field with the standard kinetic term and a time-dependent
effective  mass $m^2=z_{\tau\tau}/z$ in flat space-time. Note that
the Eq.(15) is formally the same as the case of ordinary scalar
field, but it is new because the definition of variable $z$ in
Eq.(14) is different from the one in Ref.[15]. In fact $z$ is
dependent on the equation of tachyon motion(4).Quantizing
$u(\tau,\mathbf{x})$, we have

\begin{eqnarray}
\hat{u}(\tau,
{\mathbf{x}})&=&\int\frac{d^3{\mathbf{k}}}{(2\pi)^{3/2}}\left[u_k(\tau)
\hat{a}_{{\mathbf{k}}}e^{i{\mathbf{k}}\cdot{\mathbf{x}}}+u_k^{*}(\tau)
\hat{a}_{{\mathbf{k}}}^{\dag}e^{-i{\mathbf{k}}\cdot{\mathbf{x}}}\right]\nonumber\\
\left[\hat{a}_{\mathbf{k}},
\hat{a}_{\mathbf{l}}^{\dag}\right]&=&\delta^3({\mathbf{k}}-{\mathbf{l}}),
\hat{a}_{\mathbf{k}}|0\rangle=0, etc.\hspace{3cm}
\end{eqnarray}

\noindent From Eq.(15), the mode functions $u_k$ satisfy following
equation

\begin{equation}
\frac{d^2u_k}{d\tau^2}+\left(k^2-\frac1z\frac{d^2z}{d\tau^2}\right)u_k=0\hspace{3cm}
\end{equation}

\noindent Using Eqs.(4),(7) and (8), it is easy to find that

\begin{eqnarray}
\frac{1}{2a^2H^2}\frac{1}{z}\frac{d^2z}{d\tau^2}&=&1+4\epsilon(T)-3\eta(T)+9\epsilon^2(T)\nonumber\\
&-&14\epsilon(T)\eta(T)+2\eta^2(T)+\frac12\xi^2(T)
\end{eqnarray}

\noindent where

\begin{equation}
\epsilon(T)=\frac23\left[\frac{H^{\prime}(T)}{H^2(T)}\right]^2\hspace{3cm}
\end{equation}

\begin{equation}
\eta(T)=\frac13\frac{H^{\prime\prime}(T)}{H^3(T)}\hspace{3cm}
\end{equation}

 \noindent and

\begin{equation}
\xi(T)=\frac23\left(\frac{H^{\prime}H^{\prime\prime\prime}(T)}{H^6(T)}\right)^{\frac12}\hspace{3cm}
\end{equation}

\noindent Using Eqs.(7)-(10), we can give the exact expression in
which the mode function$u_k$ is related to the field $T$. One has
the boundary conditions

\begin{equation}
u_k\rightarrow\frac1{2k}e^{-ik\tau},\hspace{0.3cm}aH\ll k
\hspace{3cm}
\end{equation}

\begin{equation}
u_k\propto z,\hspace{0.3cm}aH\gg k \hspace{3cm}
\end{equation}

\noindent which guarantees that the perturbation behaves like a
free field well inside the horizon and is fixed at superhorizon
scales. Eqs.(14)-(17) make a difference between tachyon and
ordinary scalar, because the curvature perturbations couple to the
stress-energy of tachyon field.

On each scale $\mathcal{R}$ is constant well outside the horizon.
Its spectrum is defined by

\begin{equation}
{\mathcal{R}}=\int\frac{d^3{\mathbf{k}}}{(2\pi)^{3/2}}{\mathcal{R}}_{\mathbf{k}}
(\tau)e^{i{\mathbf{k}}\cdot{\mathbf{x}}},\hspace{3cm}
\end{equation}

\begin{equation}
\langle{\mathcal{R}}_{\mathbf{k}}{\mathcal{R}}_{\mathbf{l}}^{*}\rangle=\frac{2\pi}{k^3}
{\mathcal{P_R}}\delta^3({\mathbf{k}}-{\mathbf{l}}),\hspace{3cm}
\end{equation}

\noindent where ${\mathcal{P_R}}(k)$ is the power spectrum. From
eq.(25), we have

\begin{equation}
{\mathcal{P_R}}^{\frac12}(k)=\sqrt{\frac{k^3}{2\pi^2}}\left|\frac{u_k}{z}\right|\hspace{3cm}
\end{equation}

 Furthermore, we find a simple relation
as follows

\begin{equation}
\frac{dz}{d\tau}=-a^2\left(\frac23\epsilon\right)^{\frac12}(1-2\eta+3\epsilon)\hspace{3cm}
\end{equation}

\noindent The exact inflationary solutions for the mode equations
of curvature perturbations might be found from two cases. Type I:
we start from the quantity $z$ is a constant, which is equivalent
to requiring $1-2\eta+3\epsilon=0$. Type II: we also might start
from the quantities $\epsilon(T),\eta(T)$and $\xi(T)$ are
constants.

\section{The solution of Type I}
In this case, we should demand that $H$ satisfies the differential
equation

\begin{equation}
H^4-\frac23HH^{\prime\prime}+2H^{\prime^2}=0 \hspace{3cm}
\end{equation}

\noindent which has the solution

\begin{equation}
H(T)=\frac1{\sqrt{ AT+B-\frac32T^2}}\hspace{3cm}
\end{equation}

\noindent where $A$ and $B$ are arbitrary integration constants.
However, we can chose $A=0$ without loss of generality, as it can
be recovered by making a translation of the field, $T\rightarrow
T+\frac{A}3$. From Eqs.(8)-(10), we have the corresponding tachyon
cosmology,

\begin{equation}
V(T)=\frac3{\kappa}\sqrt{\frac{B-\frac52T^2}{\left(B-\frac32T^2\right)^3}}\hspace{3cm}
\end{equation}

\begin{equation}
a(T)=\frac{a_0T_0}T\hspace{3cm} \end{equation}

\begin{eqnarray}
t(T)&&=\left(B-\frac32T^2_0\right)^{\frac12}-\left(B-\frac32T^2\right)^{\frac12}\\
&&+\frac{\sqrt{B}}2\ln\frac{\left[\sqrt{B}-\left(B-\frac32T^2_0\right)^{\frac12}\right]
\left[\sqrt{B}+\left(B-\frac32T^2\right)^{\frac12}\right]}
{\left[\sqrt{B}-\left(B-\frac32T^2\right)^{\frac12}\right]
\left[\sqrt{B}+\left(B-\frac32T^2_0\right)^{\frac12}\right]}\nonumber
\end{eqnarray}

\noindent The conformal time is

\begin{eqnarray}
\tau(T)&=&\frac1{2a_0}\left(B-\frac32T^2_0\right)^{\frac12}-\frac{T}{2a_0T_0}
\left(B-\frac32T^2\right)^{\frac12}\\
&+&\sqrt{\frac23}\frac{B}{a_0T_0}
\arcsin\sqrt{\frac3{2B}}T_0-\sqrt{\frac23}\frac{B}{a_0T_0}
\arcsin\sqrt{\frac3{2B}}T\nonumber
\end{eqnarray}

\noindent It is easy to find that $\tau$ tends to a constant value
at late time, or as $T$ goes to zero.

For this solution, the cosmological properties are easily derived.
As the tachyon field $T$ goes to zero or infinite, the potential
$V(T)$ tends to non-zero value $\frac3{\kappa B}$ or zero. The
motion is not inflationary at all time. From Eq.(6), we find that
the period of accelerated expansion corresponds to
$\dot{T}^2<\frac23$ and decelerate otherwise. Thus inflation
occurs only when $|T|<\sqrt{\frac{B}3}$. If this model was to
produce all the 50 e-foldings of inflation needed to solve the
initial conditions problems in the standard model of cosmology,
tachyon field must evolve to be close to zero.

Next, we discuss the spectrum of curvature perturbations produced
by this model, which is similar to one of the inflation model
driven by ordinary scalar field[16]. The solution of mode equation
(17) is simple,

\begin{equation}
u_k(\tau)=\frac1{\sqrt{2k}}e^{-ik\tau}\hspace{3cm}
\end{equation}

\noindent for the growing mode, after we have imposed the boundary
conditions. Since the conformal time $\tau$ tends to a constant,
mode function $u_k$ is essentially fixed at super-horizon scales.
The spectral index $n_R$ is

\begin{equation}
n_R-1 \equiv \frac{d\ln{\mathcal{P_R}}} {d\ln{k}}=2\hspace{3cm}
\end{equation}

\noindent Note that this result is exact and independent of
tachyon field $T$. By now,the inflationary universe is generally
recognized to be the most likely scenario that explains the origin
of the Big Bang. So far, its predictions of the flatness of the
universe and the almost scale-invariant power spectrum of the
curvature perturbation that seeds structure formations are in good
agreement with the cosmic microwave background(CMB) observations.
The key data of CMB are the curvature perturbation magnitude
measured by COBE[17] and its power spectrum index $n_R$[18]

\begin{equation}
\delta_H\simeq1.9\times10^{-5}\hspace{3cm}
\end{equation}

\noindent  and

\begin{equation}
|n_R-1|<0.1\hspace{3cm}
\end{equation}

\noindent Therefore, the spectrum of this model is "blue" as it
possesses more power at large values of $k$, or small scales,
which is ruled out by the observable universe. However, this model
can be used to probe the accuracy of the first and second order
approximations for the mode equation.

\section{ The solution of Type II}

In this case, we should demand that $H$ satisfies the differential
equations

\begin{equation}
\epsilon(T)=const.,\hspace{0.3cm} \eta(T)=const.\hspace{0.3cm}
{\mathrm and}\hspace{0.3cm} \xi(T)=const.\hspace{3cm}
\end{equation}

\noindent which have the solution

\begin{equation}
H(T)=\frac{\left(\frac23n\right)^{\frac12}}{T-T_0}\hspace{3cm}
\end{equation}

\noindent and

\begin{equation}
\epsilon(T)=\frac1n,\hspace{0.3cm} \eta(T)=\frac1n.\hspace{0.3cm}
{\mathrm and} \hspace{0.3cm}\xi(T)=\frac{\sqrt{6}}n\hspace{3cm}
\end{equation}

\noindent From Eqs.(8)-(10), we have the corresponding tachyon
cosmology,

\begin{equation}
V(T)=\frac2{\kappa}\left(n^2-\frac{n}3\right)^{\frac12}\left(T-T_0\right)^{-2}
\hspace{3cm}
\end{equation}

\begin{equation}
a(T)=a_0(T-T_0)^{n}\hspace{3cm}
\end{equation}

\begin{equation}
t(T)=\left(\frac32n\right)^{\frac12}(T-T_0)\hspace{3cm}
\end{equation}

\noindent The conformal time is

\begin{equation}
\tau(T)=a^{-1}_0(1-n)\left(\frac32n\right)^{\frac12}(T-T_0)^{1-n}\hspace{3cm}
\end{equation}

\noindent It is easy to  find that the condition for inflation is
$n>1$, which corresponds to $\dot{T}^2<{\frac23}$. This model
actually inflate forever. The model equation (17) is solved in
terms of Bessel functions,
$\mu_k(\tau)=(k\tau)^{\frac12}\left[C_1J_{\mu}(k\tau)+C_2J_{-\mu}(k\tau)\right]$
where $C_1$ and $C_2$ are constants fixed from Eq.(22). The
spectrum may be calculated exactly to read

\begin{equation}
{\mathcal{P_R}}(k)=\frac{\kappa}{\pi^2A^2}2^{2\mu-1}\Gamma^2(\mu)k^{-2\mu}\hspace{3cm}
\end{equation}

\noindent where $A$ is a constant

\begin{equation}
A=a^{\frac32-\mu}_0(n-1)^{\mu-\frac12}\left(\frac32n\right)^{\frac{\mu}2-\frac14}
\hspace{3cm}
\end{equation}

\noindent The corresponding spectral index is

\begin{equation}
n_R=1-\frac4{n-1}\hspace{3cm}
\end{equation}

\noindent In particular, we have $n_R\approx1$ for $n\gg1$. In
this limit, the solution and slow roll inflation with
$\epsilon=\eta=\frac1n$ agree at the leading order in $\epsilon$.
Only for $n>201$, i.e. for $1>n_R>0.98$, the error of slow roll
approximation is less than 1 percent.

We briefly conclude the main result of this paper. We find two
family of exact solutions for the mode equation (17) on curvature
perturbations of tachyon inflation. This calculation is done to
linear order, ignoring both gravity and self-interactions of the
tachyon field. Since the observed anisotropies are small, this
approximation is considerably more accurate than the slow-roll
approximation, and we need not attempt to go beyond it, though it
is possible to extend calculations beyond linear perturbation
theory[19]. These models can be probe the accuracy of the first
and second order expressions for the curvature perturbation
spectra. In fact, almost all analytical predictions for
perturbation spectra from inflation rely on the slow roll
approximation. Furthermore, the parameter $\eta$ becomes large
near the top of the tachyon potential in type I model, indicating
a breakdown of the slow roll assumption. In the cases that $n$ are
not large enough, the slow roll conditions are badly violated by
type II model. Therefore, the first order expression does not give
good agreement with the exact for all solutions. But we have
confidence that we use slow roll approximation in more realistic
situations, since the second order expression for the spectral
index of curvature perturbation can match the exact result to
within 10 percent over most of the inflationary epoch. Finally, we
point out that the slow-roll approximation is the assumption that
the field evolution is dominated by drag from the expansion is
small parameters $\epsilon$=0. We have found the solutions of mode
equation are again Hankel function of the first kind
$H_{\mu}^{(1)}(-k\tau)$ with $\mu=\frac32+4\epsilon-2\eta$.

\bigskip
\noindent\textsc{acknowledgments}

\noindent This work was partially supported by National Nature
Science Foundation of China under grant 19875016, Foundation of
Shanghai Development for Science and Technology under grant
01JC14035, Foundation of Shanghai Educational Ministry under grant
01QN86 and Foundation of Shanghai Science and Technology  Ministry
under grant 02QA14033.

\end{document}